\documentclass[aps, prl, preprint, amsmath, amssymb, showpacs, superscriptaddress]{revtex4}

\usepackage{epsfig}

\begin{document}

\title{Zero-Field Dichroism in the Solar Chromosphere}
\date{\today}
\author{R. \surname{Manso Sainz}}
\affiliation{Instituto de Astrof\'\i sica de Canarias, V\'\i a L\'actea s/n,
E-38200 La Laguna, Tenerife, Spain}
\author{J. \surname{Trujillo Bueno}}
\affiliation{Instituto de Astrof\'\i sica de Canarias, V\'\i a L\'actea s/n,
E-38200 La Laguna, Tenerife, Spain}
\affiliation{Consejo Superior de Investigaciones Cient\'\i ficas}

\begin{abstract}

We explain the linear polarization of the
Ca {\sc ii} infrared triplet observed close to the
edge of the solar disk. In particular, we demonstrate that
the physical origin of the enigmatic polarizations of the 866.2 nm and 854.2 nm lines
lies in the existence of atomic polarization in their
metastable $^2{\rm D}_{3/2\,,\,5/2}$ lower levels, which produces 
differential absorption of polarization 
components (dichroism).
To this end, we have solved the
problem of the generation and transfer
of polarized radiation by taking fully into account all the relevant
optical pumping mechanisms in multilevel atomic models. 
We argue that `zero-field' dichroism may
be of great diagnostic value in astrophysics.

\vskip 2truecm

{\bf Published in Physical Review Letters: Volume 91, Number 11, 111102, (2003)}

\vskip 5truecm

.

\end{abstract}

\pacs{95.30.-k, 95.30.Gv, 95.30.Jx, 32.80.Bx, 96.60.-j}

\maketitle

Spectropolarimetry provides key information on the physical conditions
and geometry of astrophysical plasmas otherwise unattainable 
via conventional spectroscopy \cite{TBMS02}.
The remote sensing of cosmical magnetic fields via the
Zeeman effect is the most well-known example.
In the presence of a magnetic field, the magnetic substates of 
atomic and molecular energy levels split, leading to a
frequency shift of the $\sigma^\pm$ ($\Delta M=\pm1$) 
and $\pi$ ($\Delta M=0$) transitions \cite{MZ34}.
If the Zeeman splitting is a significant fraction of the spectral line width, 
the polarization of the different components does not cancel out 
and an observable polarization pattern can then be produced 
via the emission process,
as well as through dichroism and anomalous dispersion
phenomena. 
The observed line polarization then depends on the strength and orientation
of the magnetic field vector along the line of sight.

Yet there is a more fundamental mechanism producing
linear polarization in spectral lines.
In the outer layers of stellar atmospheres, where light escapes  
through the stellar surface, the atoms and molecules are illuminated 
by an anisotropic radiation field.
The ensuing radiation pumping produces 
population imbalances among the magnetic sublevels of energy levels
(that is, atomic polarization), in such a way that the population of substates
with different values of $|M|$ are different. 
This is termed {\em atomic level alignment}.
As a result, the emission process can generate linear polarization
in spectral lines without the need for a magnetic field. This is known as 
scattering line polarization \cite{MZ34}. However,
light polarization components {\em will also be differentially absorbed 
when the lower level of the transition is polarized} \cite{TBL97, TB99}.
Thus, the medium becomes dichroic simply because the light itself 
is escaping from it. 
Since this process has nothing to do with magnetic fields, 
we call it {\em zero-field dichroism}. This mechanism may be
of great diagnostic value because in
weakly magnetized astrophysical plasmas, like the `quiet'
solar atmosphere, the contribution of the transverse
Zeeman effect to the linear polarization is normally negligible.
In order for zero-field dichroism to be operative, 
the atomic polarization 
of the lower levels must survive depolarizing effects, such as those
produced by elastic collisions. Moreover,
weak magnetic fields modify the atomic level polarization via the 
Hanle effect \cite{hanle, TB01}, 
where the magnetic field strength $B$ (in gauss) that produces
a significant change is 
$B \approx 1.137\times10^{-7} / (t_{\rm life} g_{{}_L})$
($t_{\rm life}$ and $g_{{}_L}$ being the level lifetime and Land\'e factor, 
respectively).
Since the lifetimes of the lower levels of the transitions of interest are usually much larger
than those of the upper levels, it is clear that diagnostic techniques
based on this mechanism are sensitive to relatively low densities and
magnetic field strengths.

In this letter, we demonstrate that zero-field dichroism is the key mechanism 
underlying the linear polarization of the   
Ca {\sc ii} infrared triplet detected in observations
close to the edge of the solar disk,  
in very quiet regions far away from sunspots.
The observed fractional 
polarization amplitudes of the
Ca {\sc ii} lines at 849.8, 854.2 and 866.2 nm are 
$Q/I\approx$ 0.035\%, 0.13\% and 0.12\%, 
respectively \cite{SKG00},
the polarization signal in the 866.2 nm line being especially intriguing.
This is a ${}^2{\rm D}_{3/2}\rightarrow {}^2{\rm P}_{1/2}$ transition
whose upper level (${}^2{\rm P}_{1/2}$) cannot be aligned.
Therefore, emission following anisotropic radiative
excitation cannot produce linear polarization in this spectral line 
and its detection has been considered `enigmatic' \cite{SKG00}.

\begin{figure}
\centerline{\epsfig{figure=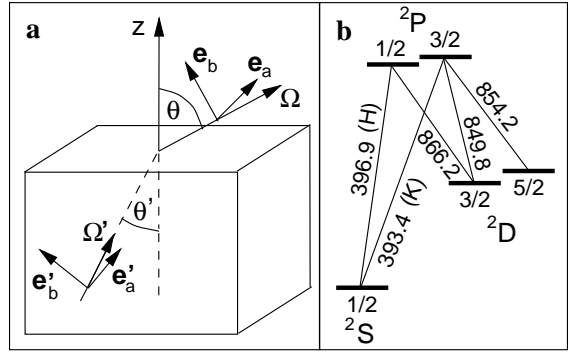, width=7.5cm}}
\caption{(a) Scattering geometry. 
Ray propagating along the direction $\boldsymbol{\Omega}$ 
with polar angle $\theta$. The
reference polarization direction for $Q>0$ is along
$\boldsymbol{{\rm e}}_a$, i.e., parallel to the stellar surface.
(b) Grotrian diagram of our atomic model for Ca {\sc ii}
indicating the total angular momentum of the levels and the spectral
line wavelengths in nm.}
\end{figure}

Consider a plane-parallel stellar atmosphere
without magnetic fields. 
For reasons of symmetry, choosing the reference polarization directions
as in Fig.~1, % (i.e., parallel and perpendicular to the stellar surface), 
the polarization state of a radiation beam can be described in terms of
the intensity $I$ and the Stokes parameter $Q$ for linear polarization. 
The transfer of $I$ and $Q$ along a ray path element d$s$ is described
by the following equations:
\begin{align} 
\begin{split}
\frac{\rm d}{{\rm d}s}\,I &=\,\epsilon_I\,\,-\,\eta_II\,-\eta_QQ\;,
\label{1}
\end{split}
\displaybreak[0] \\
\begin{split}
\frac{\rm d}{{\rm d}s}\,Q &=\,\epsilon_Q\,-\,\eta_QI\,-\eta_IQ\;,
\label{2}
\end{split}
\end{align} 
where $\epsilon_I$ and $\epsilon_Q$ are the corresponding emissivities,
while $\eta_I$ and $\eta_Q$ are the absorption and
dichroism coefficients, respectively.
In a spectral line, the emissivities 
depend on 
the excitation state of the upper level
of the transition. Conversely, $\eta_I$ and $\eta_Q$ are given by %in terms of 
the excitation state of the transition's lower level.
We have considered an additional contribution 
to $\epsilon_I$ and $\eta_I$ from a %an unpolarized
background continuum assumed to be unpolarized (i.e., $\eta_I^{\rm cont}$ and 
$\epsilon_I^{\rm cont} = \eta_I^{\rm cont}B_\nu$, $B_\nu$ being the Planck 
function), which is a very good approximation 
towards the red part of the solar spectrum (see \cite{G00}).
For simplicity, 
we neglect stimulated emissions \footnote{For a blackbody at the effective
temperature of the Sun ($T\approx 5800$ K), the number of photons per
mode even at 900 nm is $\bar{n}\approx 6\times 10^{-2} \ll 1$, and stimulated
emission can be safely neglected.}.

We choose the quantization axis for angular momentum along the Z-axis 
of Fig.~1.
Since the radiation field 
is
axially symmetric around this direction, 
no quantum interferences are induced between magnetic sublevels
and the excitation state of each atomic level is fully described 
by the individual populations (occupation probabilities) $N_M$ 
of each of its magnetic sublevels.
Equivalently, we can describe the excitation state 
of each level of total angular momentum, $J$, by means of
the following components of the atomic density matrix \cite{O77},
\begin{equation}
{\rho^K_0}(J)=\sum_{MM'}(-1)^{J-M}{\sqrt{2K+1}}
\left( \begin{array}{ccc}
J&J&K \\
M&-M^{'}&0
\end{array} \right)N_M, \quad K=0,1,2,...,2J,
\end{equation}
which are just linear combinations of $N_M$.
Because the radiation field in the assumed model atmosphere has
no net circular polarization, we have an internal 
symmetry in our problem: $N_{M}=N_{-M}$; therefore, 
only components $\rho^K_0$ with $K$ even can be different from zero.
Of particular interest is $\rho^0_0$, which is
$(2J+1)^{-1/2}$ times the overall population of the level, and
the alignment coefficient $\rho^2_0$, which quantifies the population
imbalance between the magnetic sublevels.
The line emissivities can be expressed as \cite{L84, TBL97}
\begin{align}
\epsilon_I^{\rm line}&={\epsilon_{0}}\rho^0_0+ 
{\epsilon_{0}}w_{J_u J_\ell}\frac{1}{2\sqrt{2}}(3\mu^2-1)\rho^2_0, 
\label{3} 
\displaybreak[0] \\
\epsilon_Q^{\rm line}&= {\epsilon_{0}}w_{J_u J_\ell}\frac{3}{2\sqrt{2}}
	(1-\mu^2)\rho^2_0, 
\label{4}
\end{align}
where the $\rho^0_0$ and $\rho^2_0$ values are those of the 
upper level of the line transition under consideration, 
$\mu=\cos\theta$ (see Fig.~1),
and $\epsilon_0=(h\nu/4\pi)A_{u\ell}{\phi_\nu}{\cal N}\sqrt{2J_u+1}$
(where $\nu$ is the line frequency,
$A_{u\ell}$ is the Einstein coefficient for the spontaneous emission process, 
$\cal N$ is the total number of atoms per unit volume,
and $\phi_\nu$ is the absorption profile).
$w_{J_u J_\ell}$ is a numerical coefficient whose value depends
on the total angular momentum of the upper and lower levels, 
taking values $-2\sqrt{2}/5$, $\sqrt{2}/10$, and 0 for the 
lines at 849.8, 854.2, and 866.2 nm, 
respectively.
On the other hand, $\eta_I$ and $\eta_Q$
are given by identical expressions, but using the $\rho^0_0$ and $\rho^2_0$
values of the lower level,
$\eta_0=(h\nu/4\pi)B_{\ell u}{\phi_x}{\cal N}\sqrt{2J_\ell+1}$ instead of $\epsilon_0$
(with $B_{\ell u}$ the Einstein coefficient for the absorption process), and
$w_{J_\ell J_u}$ instead of $w_{J_u J_\ell}$, where the $w_{J_\ell J_u}$ values are
$-2\sqrt{2}/5$, $\sqrt{7}/5$ and $\sqrt{2}/2$ for the aforementioned 
lines, respectively.

It can be deduced 
from Eqs.~(\ref{1})-(\ref{2}) and the expressions of $\epsilon_{I, Q}$ and
$\eta_{I, Q}$, that the emergent fractional linear polarization 
at the core of a strong spectral line is 
approximately given by \cite{TB99, TB01}:
\begin{equation}
Q/I\,\approx\,{\frac{3}{2\sqrt{2}}}(1-\mu^2)
[{w_{J_u J_\ell}}\,\sigma^2_0({J_u})\,-\,{w_{J_\ell J_u}}\,\sigma^2_0({J_l})],
\label{5}
\end{equation}
where $\sigma^2_0=\rho^2_0/\rho^0_0$ is the fractional atomic alignment of
the level under consideration. 
In this formula the $\sigma^2_0$ values are those at 
optical depth ${\tau}={\int{\eta_I}{\rm d}s}$ 
along the line of sight, where $\tau\approx 1$.
Eq.~(\ref{5}) shows clearly that
the observed fractional polarization in a given spectral line has in general
two contributions, one from the atomic polarization of the upper level
caused exclusively by emission 
from a polarized upper level
($\sigma^2_0(J_u)$), and another one from the atomic polarization 
of the lower level ($\sigma^2_0(J_l)$). 
As shown here, the latter one
(caused by the selective absorption resulting from the 
population imbalances of the lower level) plays
a key role in producing the linear
polarization pattern of the Ca {\sc ii} IR triplet. 

The atomic model in Fig.~1 is sufficiently realistic for modeling 
the Ca {\sc ii} infrared triplet.
Since calcium has no hyperfine structure, the  
only levels that may carry atomic alignment are
the upper level $^2{\rm P}_{3/2}$
and the two metastable levels
$^2{\rm D}_{3/2, 5/2}$.
Therefore, nine density
matrix elements are required to describe the atomic excitation: 
the overall population of each of the
five levels (i.e., $\sqrt{2J+1}\rho^0_0(J)$), the atomic alignment of the
two levels with $J=3/2$, and the $\rho^2_0$ and $\rho^4_0$ values corresponding
to the level with $J=5/2$.
These irreducible tensorial components of the atomic density matrix
are governed by the statistical equilibrium equations,
$0={\rm d}\rho^K_Q(i)/{\rm d}t=
-\sum_{K'} R_{iKK'}\rho^{K'}_0(i)
+\sum_{jK'} T_{jiKK'}\rho^{K'}_0(j)$, where
the relaxation matrix has the three contributions 
$R_{iKK'}=\sum_\ell R_{i\ell KK'}+\sum_u R_{iuKK'}+D^K$
caused by decays toward lower levels, excitations towards upper levels,
and depolarization by elastic collisions, respectively, while the transfer matrix
has contributions caused by decays from upper levels ($T_{uiKK'}$)
and excitations from lower levels ($T_{\ell iKK'}$) \cite{L84}.
The relaxation rates read
\begin{align}
R_{i\ell KK'}&=\delta_{KK'}\,A_{i\ell}\,+\,\delta_{KK'}\,C_{i\ell},
\label{7}
\displaybreak[0] \\
R_{iuKK'} &= \delta_{KK'}\, B_{iu}J^0_0 \,+\,
	s_{KK'}\,B_{iu}J^2_0 \,+\,\delta_{KK'}\, C_{iu},
\label{8}
\end{align}
and the transfer rates are
\begin{align}
T_{uiKK'}&=\sqrt{\frac{2J_u+1}{2J_i+1}} 
	[\delta_{KK'}\,p_{K'}\,A_{ui}\,+\,\delta_{KK'} \,
	p_{K'}\, C_{ui}],
\label{9}
\displaybreak[0] \\
T_{\ell iKK'} &= \sqrt{\frac{2J_\ell+1}{2J_i+1}} 
	[\delta_{KK'}\,p_{K'}\, B_{\ell i}J^0_0 \,+\,
	q_{KK'}\,B_{\ell i}J^2_0 \,+\, \delta_{KK'} \,
	p_{K'}\, C_{\ell i}],
\label{10}
\end{align}
where 
$p$, $q$, and $s$ are numerical coefficients that depend only
on the quantum numbers of the levels involved in the transition
(see Table~1). Since the radiation field is axisymmetric
the absorption rates depend only on the line-integrated mean intensity
($J^0_0$), and on
\begin{align}
{J}^2_0&=\,\frac{1}{4\sqrt{2}}\int {\rm d}\nu\,\int_{-1}^1 {\rm d}\mu'\,\phi_\nu
	[(3\mu'^2-1)I+3(1-\mu'^2)Q].
\label{12}
\nonumber
\end{align} 
It is important to note that in stellar atmospheres the polarization degree is small ($Q/I\ll 1$),
and the main contribution to $J^2_0$ results from the 
intensity average. 
We call ${\cal A}={J}^2_0/{J}^0_0$ the `degree of anisotropy' of the radiation
field. It determines
the fractional atomic polarization that can be induced by optical pumping processes.
In the solar atmosphere ${\cal A}\ll 1$, hence 
$\sigma^2_0\ll 1$. 
Collisional rates $C_{ij}$ due to collisions with an isotropic Maxwellian
distribution of electrons are modeled in analogy to the excitation
by an isotropic unpolarized radiation field.
We have used the inelastic collisional rates for Ca {\sc ii} tabulated in
\cite{SL74}, while the depolarizing rates $D^{(K)}$ due to 
collisions with neutral hydrogen atoms
have been estimated according to Ref.~\cite{LtH71}.

\begin{table}
\caption{Numerical coefficients of Eqs.~(\ref{7})-(\ref{10})
calculated for the 5-level Ca {\sc ii} model of Fig. 1.
$p_0\equiv1$ for all transitions. 
For the 396.9 nm line ($J_l=J_u=1/2$) any other coefficients are zero.}
\begin{tabular}{cccccccc}
\hline \hline 
line & $p_2$ & $q_{20}$ & $q_{22}$ & $s_{20}$ &  $s_{22}$ &  $s_{42}$ &  $s_{44}$ \\
\hline
393.4 & --- & $1/\sqrt{2}$ & --- & --- & --- & --- & --- \\
866.2 & --- & --- & --- & $1/\sqrt{2}$ & --- & --- & --- \\
849.8 & $1/5$ & $-2\sqrt{2}/5$ & $-2\sqrt{2}/5$ & $-2\sqrt{2}/5$ & --- & --- & --- \\
854.2 & $\sqrt{14}/5$ & $\sqrt{2}/10$ & $2/(5\sqrt{7})$ & $\sqrt{7}/5$ &
	$\sqrt{2}/7$ & $9\sqrt{6}/70$ & $-\sqrt{2}/7$ \\
\hline \hline
\end{tabular}
\end{table}

\begin{figure}
\centerline{\epsfig{figure=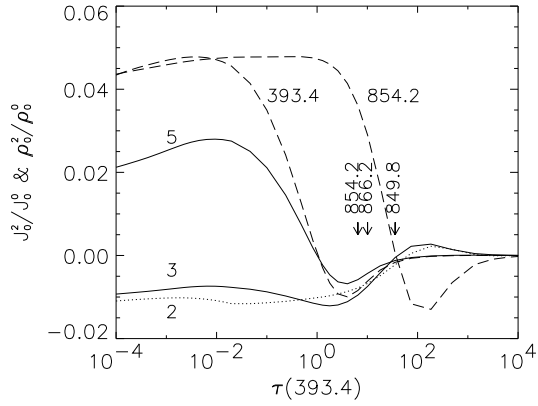, width=7.5cm}}
\caption{Variation of $\rho^2_0/\rho^0_0$
	in 2nd, 3rd and 5th levels of our Ca {\sc ii} model 
	as a function of the K-line
	optical depth calculated along the vertical direction 
	(solid and dotted lines).
	Dashed lines show $J^2_0/J^0_0$ in
	the K and 854.2 nm lines.
	Arrows indicate the layer in the FAL-C model 
	atmosphere \cite{FAL91} where, for each of the three IR lines,
	$\tau=1$ along the $\mu=0.1$ line of sight.
	\label{fig04}}
\end{figure}

\begin{figure}
\centerline{\epsfig{figure=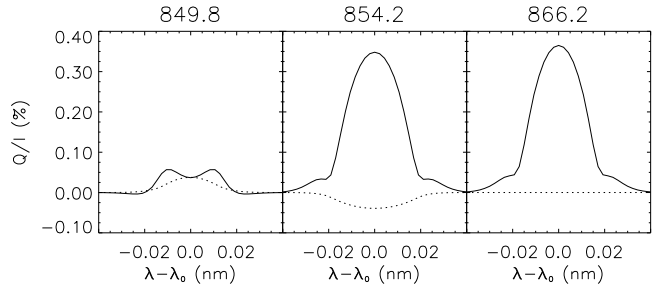, width=9cm}}
\caption{Emergent fractional polarization
	of the Ca {\sc ii} infrared triplet
	at $\mu=0.1$ in the
	FAL-C semi-empirical model of the solar atmosphere (solid lines).
	The shapes of the calculated $Q/I$ profiles 
	and their relative line-core amplitudes
	are in good agreement
	with the reported observations (see \cite{SKG00}).
	Dotted lines: $Q/I$ profiles if
	the {\em long-lived} 
	metastable $^2{\rm D}_{3/2\,,\,5/2}$ levels were completely unpolarized.
	\label{fig02}}
\end{figure}

Figure~\ref{fig04} shows results of the self-consistent solution of the radiative 
transfer and statistical equilibrium equations in a semi-empirical model
\cite{FAL91} of the solar atmosphere. The numerical solution has been obtained applying the 
iterative methods outlined in Ref.~\cite{TB99}. The figure
indicates how the degree of anisotropy corresponding
to the K-line and to the 854.2 nm line vary with
the K-line optical depth. 
These spectral lines are representative of the
radiation fields in the UV doublet and IR triplet, respectively.
The figure shows also the behavior of the fractional atomic
alignments $\sigma^2_0$ of the 2nd, 3rd and 5th levels.
Note that they have sizable values, even at the depths in the stellar atmosphere model
where the line optical depths are unity along the line of sight.

The atomic alignment of the 5th level is governed by the radiation field
in the K-line, with optical pumping in
the lines at 849.8 nm and 854.2 nm playing only a marginal role.
In fact, the height variation of $\sigma^2_0(5)$ closely follows that 
of the degree of anisotropy in the K-line throughout 
the whole atmosphere, with $\sigma^2_0(5)\approx
\sqrt{2}/2\,(J^2_0/J^0_0)({\rm K})$. 
On the other hand, there are two optical pumping processes able to introduce
atomic alignment in the lower levels of the Ca {\sc ii} IR triplet
(the metastable levels):
{\em repopulation pumping}, which
results from the spontaneous decay of the polarized 5th level;
{\em depopulation pumping}, which occurs when some lower 
state sublevels absorb anisotropic light more strongly than others.  
Since the infrared triplet lines are much weaker than the ultraviolet H- and K-lines
($\eta_I^{\rm line}/\eta_I^{\rm cont}\approx 10^5$ for the former, while 
$\eta_I^{\rm line}/\eta_I^{\rm cont}\approx 10^7$ 
for the latter lines), 
the IR line photons escape from deeper atmospheric regions,
where the $J^2_0/J^0_0$ values of
the IR line transitions are much larger than the (negligible)
$\sigma^2_0$ value of the 5th level at such depths (see Fig. \ref{fig04}).
Consequently, at such atmospheric depths the long-lived metastable levels 2 and 3 
are mainly aligned by {\em depopulation pumping} rather than
by spontaneous emission from the upper level with $J=3/2$.

The solid lines in Fig. \ref{fig02} show the emergent fractional polarization 
of the Ca {\sc ii} infrared triplet calculated at $\mu=0.1$.
Dotted lines show the emergent fractional polarizations if 
the metastable levels were unpolarized
and dichroism negligible.
Since the upper level of the 866.2 nm line cannot be aligned, 
the $Q/I$ signal in this transition is exclusively caused by 
dichroism [see  Eq.~(\ref{5})]. 
The linear polarization observed in the 854.2 nm line can, in principle,  
be caused by both, the emission events from the polarized upper level and
to the differential absorption by the polarized lower level.
However, $|\sigma^2_0({}^2{\rm D}_{5/2})|>|\sigma^2_0({}^2{\rm P}_{3/2})|$
at $\tau(854.2)\approx 1$ (see Fig.~\ref{fig04}), while
$w_{J_\ell J_u} > w_{J_u J_\ell}$ in Eq.~(\ref{5}). 
Therefore, the observed $Q/I$ signal is {\em de facto} 
also mainly caused by dichroism in the solar atmosphere.
The 849.8 nm line forms significantly deeper in the solar atmosphere, 
where the fractional alignment of its 
upper and lower levels are close to zero,
hence its much lower polarization signal, whose sign turns out to be very
sensitive to the assumed atmospheric model. 

Note that the calculated $Q/I$ amplitudes of the 
854.2 nm and 866.2 nm lines (see Fig.~\ref{fig02})
are roughly twice the values reported in \cite{SKG00}.
This can be due either to a simplified atmospheric modeling 
(the highly inhomogeneous solar atmosphere may not be well
represented by a one-dimensional semi-empirical model),
and/or to the existence of depolarizing mechanisms.
Either case emphasizes the great diagnostic potential 
of the scattering polarization observed in the Ca {\sc ii} IR triplet. 
For example, collisional depolarizing rates five times larger 
than the nominal values used here only slightly affect 
the ensuing $Q/I$ amplitudes. 
On the other hand, in the presence  of a microturbulent magnetic field 
with a strength as low as 0.01 gauss, the atomic polarization of the 
metastable $^2{\rm D}_{3/2\,,\,5/2}$ levels are significantly reduced
and hence the emergent linear polarization
in the 854.2 nm and 866.2 nm lines 
decreases to values similar to those observed (Hanle effect).

Zero-field dichroism may also be operating 
in other astrophysical objects (e.g., accreting systems)
and should be fully taken into account
when interpreting spectropolarimetric observations in other spectral
lines besides the Ca {\sc ii} IR triplet itself, 
whose polarization has been observed recently in supernovae \cite{SN}.

Research partially funded by the Spanish Ministerio de Ciencia
y Tecnolog\'\i a through project AYA2001-1649.


\begin{thebibliography}{}
\bibitem{TBMS02} Reviews introducing the interest of 
	spectropolarimetry 
	in several fields of astrophysics appear in 
	{\em Astrophysical Spectropolarimetry}, proceedings of the XII 
	Canary Islands Winter School of Astrophysics, Puerto de la Cruz, 
	Tenerife, Spain, 2000, edited by J. Trujillo Bueno, 
	F. Moreno-Insertis, and F. S\'anchez
	(Cambridge University Press, Cambridge, U. K., 2002).
\bibitem{MZ34} J. O. Stenflo, {\em Solar Magnetic Fields: 
	Polarized Radiation Diagnostics} 
       	(Kluwer Academic Publishers, Dordrecht, 1994).	
\bibitem{TBL97} J. Trujillo Bueno and E. Landi Degl'Innocenti, 
	Astrophys. J. Lett. {\bf 482}, 183 (1997).
\bibitem{TB99} J. Trujillo Bueno, in {\em Solar Polarization}, proceedings
       	of an international workshop held in Bangalore, Bangalore, 1998,
       	edited by K. N. Nagendra \& J. O. Stenflo  
	(Kluwer Academic Publishers, Dordrecht, 1999), p. 73.		
\bibitem{hanle} W. Hanle, Z. Phys. {\bf 30}, 93 (1924).
\bibitem{TB01} J. Trujillo Bueno, in Astron. Soc. Pacific Conf. Ser. {\bf 236},
	{\em Advanced Solar Polarimetry. Theory, Observation and 
	Instrumentation}, proceedings of the 20th NSO/Sacramento Peak Summer 
	Workshop, Sunspot, New Mexico,
	2000, edited by M. Sigwarth 
	(Astron. Soc. Pacific, San Francisco, 2001), p. 161.
\bibitem{SKG00} J. O. Stenflo, C. U. Keller, and A. Gandorfer, 
	Astron. Astrophys. {\bf 355}, 789 (2000).
\bibitem{G00} A. Gandorfer, {\em The Second Solar Spectrum} 
	(vdf Hochschulverlag AG an der ETH, Z\"urich, 2000), Vol. 1.
\bibitem{O77} A. Omont, Prog. Quantum Electron. {\bf 5}, 69 (1977).
\bibitem{L84} E. Landi Degl'Innocenti, Sol. Phys. {\bf 85}, 3 (1983);
	{\bf 91}, 1 (1984).
\bibitem{SL74} R. A. Shine and J. L. Linsky, Sol. Phys. {\bf 39}, 49 (1974).
\bibitem{LtH71} F. K. Lamb and D. ter Haar, Phys. Rep. {\bf 2C(4)}, 253 (1971).
\bibitem{FAL91} J. M. Fontenla, E. H. Avrett, and R. Loeser, Astrophys. J. 
	{\bf 377}, 712 (1991).
\bibitem{SN} D. C. Leonard, A. V. Filippenko, R. Chornock, and R. J. Foley,
	Publ. Astron. Soc. Pac. {\bf 114}, 1333 (2002);
	K. S. Kawabata {\em et al.}, Astrophys. J. Lett. {\bf 580}, 39 (2002);
	L. Wang {\em et al.}, Astrophys. J. (in press) 
	[preprint arXiv:astro-ph/0303397, 2003].
\end{thebibliography}
\end{document}